\documentclass[iop]{emulateapj}

\usepackage{apjfonts}
\usepackage{hyperref}
\usepackage{graphicx}
\usepackage{bigstrut}

\newcommand{\gx}{GX 13+1}
\newcommand{\xte}{{\it RXTE}}
\newcommand{\cha}{{\it Chandra}}
\newcommand{\xmm}{{\it XMM-Newton}}
\newcommand{\vel}{km\,s$^{-1}$}
\newcommand{\flux}{erg\,s$^{-1}$\,cm$^{-2}$}

\newcommand{\nh}{$N_{\rm H}$}

\newcommand{\ledd}{$L_{\rm Edd}$}

\shorttitle{Simultaneous jets and disk winds}
\shortauthors{Homan et al.}

\begin{document}

\title{Evidence for simultaneous jets and disk winds in luminous low-mass X-ray binaries}

\author{Jeroen Homan\altaffilmark{1, 2}}
\author{Joseph Neilsen\altaffilmark{1,6}}
\author{Jessamyn L.\ Allen\altaffilmark{1,3}}
\author{Deepto Chakrabarty\altaffilmark{1, 3}}
\author{Rob Fender\altaffilmark{4}}
\author{Joel K.\ Fridriksson\altaffilmark{5}}
\author{Ronald A.\ Remillard\altaffilmark{1}}
\author{Norbert Schulz\altaffilmark{1}}

\altaffiltext{1}{MIT Kavli Institute for Astrophysics and Space Research, 77 Massachusetts Avenue 37-582D, Cambridge, MA 02139, USA; jeroen@space.mit.edu}

\altaffiltext{2}{SRON, Netherlands Institute for Space Research, Sorbonnelaan 2, 3584 CA Utrecht, The Netherlands}

\altaffiltext{3}{Department of Physics, Massachusetts Institute of Technology, 77 Massachusetts Avenue, Cambridge, MA 02139, USA}

\altaffiltext{4}{Astrophysics, Department of Physics, University of Oxford, Keble Road, Oxford OX1 3RH, UK}

\altaffiltext{5}{Anton Pannekoek Institute, University of Amsterdam, Postbus 94249, 1090 GE Amsterdam, The Netherlands}

\altaffiltext{6}{Hubble Postdoctoral Fellow}

\begin{abstract}

Recent work on jets and disk winds in low-mass X-ray binaries (LMXBs) suggests that they are to a large extent mutually exclusive, with jets observed in spectrally hard states and disk winds observed in spectrally soft states. In this paper we use existing literature on jets and disk winds in the luminous neutron star (NS) LMXB \gx, in combination with archival {\it Rossi X-ray Timing Explorer} data, to show that this source is likely able to produce jets and disk winds simultaneously. We find that jets and disk winds occur in the same location on the source's track in its X-ray color-color diagram. A further study of literature on other luminous LMXBs reveals that this behavior is more common, with indications for simultaneous jets and disk winds in the black hole LMXBs V404 Cyg and GRS 1915+105 and the NS LMXBs Sco X-1 and Cir X-1. For the three sources for which we have the necessary spectral information, we find that the simultaneous jets/winds all occur in their spectrally hardest states. Our findings indicate that in LMXBs with luminosities above a few tens of percent of the Eddington luminosity, jets and disk winds are not mutually exclusive, and that the presence of disk winds does not necessarily result in jet suppression. 
\end{abstract}


\section{Introduction}\label{sec:intro}

The accretion flows in low-mass X-ray binaries (LMXBs) are commonly accompanied by jet outflows or disk winds. The presence of jet outflows is often deduced from radio emission. Radio observations of jet outflows have revealed that they are highly collimated, often relativistic, and are launched close to the compact object \citep{fogebr2001a,fogebr2001b,fe2006,fega2014}. Disk winds are typically studied via narrow line features in X-ray spectra; they have a more equatorial geometry, are much slower, and are formed further out \citep{ne2013,dibo2015}. Observed behavior in the black hole (BH) LMXB H1743--332 suggested that jets and disk winds are causally related  \citep{miraho2006}; later, \citet{nele2009} found that in the BH LMXB GRS 1915+105 the jet outflows are typically quenched when strong disk winds are present \citep[see also][]{mirare2008}. A large study of BH LMXBs by \citet{pofebe2012} revealed that disk winds (in high-inclination systems) are mainly found in the spectrally soft X-ray state. Combined with the fact that radio jets in BH LMXBs are predominantly found in the spectrally hard X-ray state \citep{fehobe2009}, this suggested that jets and disk winds are to a large extent mutually exclusive. 

\citet{pomufe2014,pobimu2015} analyzed the data of two high-inclination neutron star (NS) LMXBs and found that the Fe K absorption lines, which are often used to trace disk winds (or disk atmospheres), showed the same dependence on spectral state as in the BH systems. However,  the spectral resolution of their data was too low to measure possible outflow velocities.

The above findings on the state dependence of disk winds in BH and NS LMXBs come mostly from systems with luminosities less than a few tens of percent of the Eddington luminosity (\ledd).  The only exception to the overall trend seen in the sample of \citet{pofebe2012} is GRS 1915+105, which has shown disk winds in a luminous hard state ($\sim$0.3 \ledd); in that observation GRS 1915+105 produced a jet and disk wind at the same time  \citep{lerere2002,nele2009}. This behavior could indicate that, as the luminosity starts approaching \ledd, the relation between disk winds and jets (and spectral state) may change from that at lower luminosities.   In this paper we explore this idea further by using existing literature on disk winds and radio jets in the NS LMXB \gx, in combination with archival {\it Rossi X-ray Timing Explorer} (\xte) data of that source.

\subsection{\gx}

\gx\ is a persistently bright NS LMXB. The source was initially classified as a so-called atoll source by \citet{hava1989} on the basis of {\it EXOSAT} data, although \citet{schatr1989} had grouped it with the Z sources, a subclass of the NS LMXBs with luminosities near or above \ledd. A recent analysis by \citet[hereafter F15]{frhore2015} showed that, at least during the lifetime of \xte\ (1995--2012), \gx\ showed properties more similar to the Z sources.

Periodic (24.27 day) absorption dips have been observed in the X-ray light curves of \gx\ \citep{dairdi2014,iadibu2014}, implying a high inclination (65$^\circ$--70$^\circ$). As in other high-inclination systems, evidence for a warm absorber was found in GX 13+1, with {\it ASCA} and \xmm\ \citep{ueasya2001,sipaoo2002}. Later observations with \cha\ revealed that the absorption lines from this warm absorber were significantly blueshifted, indicating the presence of a disk wind with outflow velocities of up to  $\sim$800 \vel\   \citep{uemuya2004,majodi2014}. Evidence for a disk wind was also seen in near-infrared spectra of \gx; these revealed a strong P Cygni profile in the Br$\gamma$ line from which an outflow velocity of $\sim$2400 \vel\ was inferred \citep{bashch1999}. It is not clear whether the winds observed in X-rays and near-infrared are one and the same, or if they represent two distinct outflows originating in different parts of the accretion disk.

\gx\  is variable in the radio \citep{gagrmo1988,howiru2004}, with luminosities similar to those of the other Z sources \citep{fehe2000}, indicating the presence of strong  jet outflows. As in most of the other Z sources \citep[see, e.g.,][]{mife2006,mimife2007,spruba2013}, the evolution of the radio flux of \gx\ suggests that the jet is strongest on the so-called normal (NB) and horizontal branches (HB) of its Z track, and strongly suppressed on its flaring branch \citep[FB,][]{howiru2004}. 

While the X-ray spectral state dependence of the radio emission in \gx\ has been studied, this is not the case for the disk wind. In this paper we show that disk winds in \gx\ are found in the same spectral state in which the radio flux reaches its maximum, suggesting that disk winds and radio jets can be produced simultaneously in this source. In Section \ref{sec:obs}  we present the \xte\ data that were used to obtain information on the X-ray spectral state of \gx\ and  summarize the literature on disk winds and jet outflows in the source. In Section \ref{sec:results} we present our results on \gx\ and investigate other LMXBs with near-\ledd\ luminosities for possible indications of simultaneous jets and disk winds. Our results are discussed in further detail in Section \ref{sec:disc}. A detailed analysis of  how the wind properties change along the Z track of \gx\ will be presented in J.\ L.\ Allen et al. (2016, in preparation).

\section{X-ray and Radio Observations of \gx}\label{sec:obs}

\subsection{\xte}

To investigate in which spectral states of \gx\  jets and disk winds were observed, we use \xte\ \citep{brrosw1993} data of the source. F15 analyzed the entire set of observations of \gx\ made with the Proportional Counter Array (PCA; \citealt{jamara2006}) on board {\it RXTE}. This data set consists of 92 individual ObsIDs. Here we make use of the color--color diagrams (CDs) produced by F15 and we refer to their work for the details of the data analysis. In Table \ref{tab:obs} we list the ObsIDs of the \xte\ observations that were carried out simultaneously with the \cha\ and Very Large Array (VLA) observations discussed in this work.

\begin{figure*}[t] 
\centerline{\includegraphics[width=15.6cm]{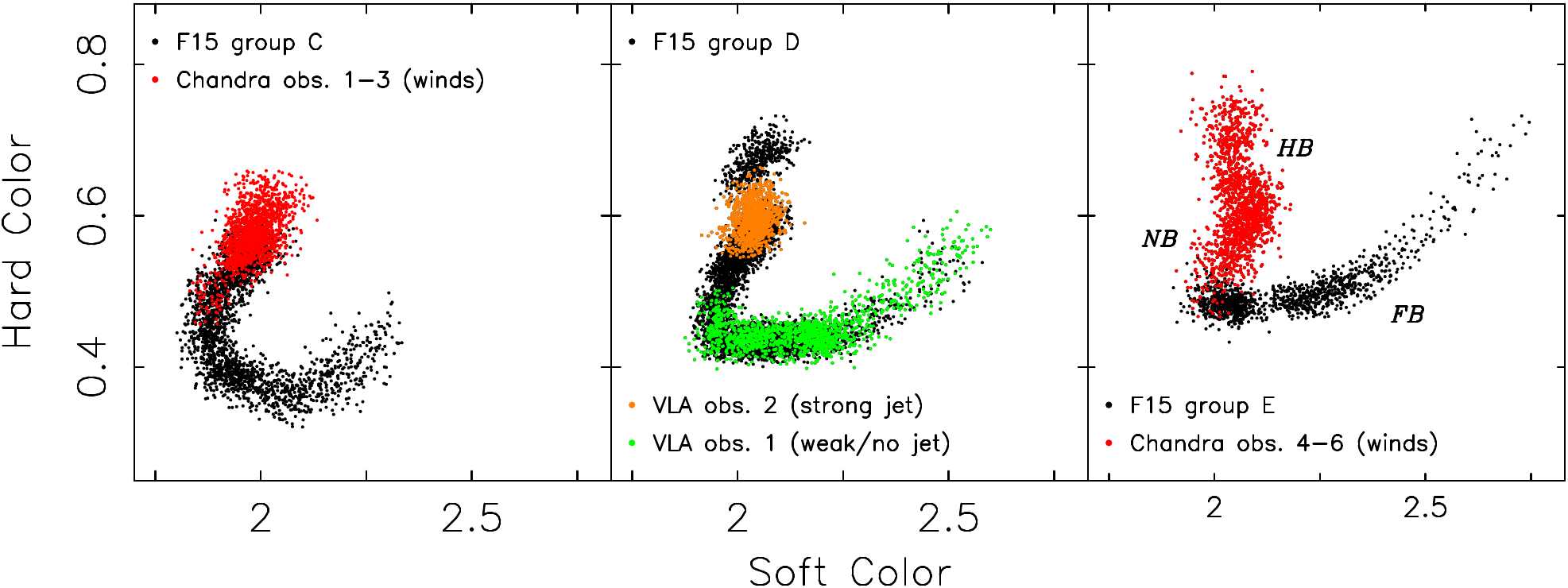}}
\caption{\xte\ CDs of \gx. The three CDs represent different stages of the secular evolution of \gx's Z track and are based on the six representative CD tracks shown in Figure 12 of F15. The data corresponding to the  \cha\ and VLA observations are shown in red and orange/green, respectively. The colored data were already part of the tracks created by F15, with the exception of the left panel, for which we found group C from F15 to offer the best match.      Indications for strong jets  (orange; middle panel) and winds (red; left and right panels) are both seen on the HB/NB.  Soft color is defined as the net counts in the 4.0--7.3 keV band divided by those in the 2.4--4.0 keV band, while hard color is defined as net counts in the 9.8--18.2 keV band divided by those in the 7.3--9.8 keV band.  See F15 for further details.} 
\label{fig:cd}
\end{figure*}

As discussed in detail by F15, \gx\ shows substantial secular evolution of its Z track in CDs and hardness--intensity diagrams (HIDs), with the shape and position of the Z track changing on timescales of days or longer (see also \citealt{screva2003}). F15 combined data into six different groups to illustrate the evolution of the Z tracks of GX 13+1 (see their Figure 12 and Table 5). In Figure \ref{fig:cd} we show three of those groups in a CD. Clear changes in the shape and position of the Z track can be seen. The three Z source branches are labeled in the right panel. The location of the disk wind and radio jet detections in the CD will be discussed in more detail in Sections \ref{sec:cha} and \ref{sec:vla}.

\begin{table}
\caption{Log of \xte\ observations}
\begin{center}
\begin{tabular}{ccc}
\hline
\hline
\xte\ ObsID & Start Date (UTC) & Simultaneous Obs.\ \\
\hline

40022-01-01-00	&	1999-07-31 23:52	&	VLA obs.\ 1	\\
40022-01-01-01	&	1999-08-01 07:23	&	$\cdots$	\\
\hline
40022-01-02-000    &      1999-08-04 02:45   &     VLA obs.\ 2 \\
40022-01-02-00	&	1999-08-04 10:39	&	$\cdots$	\\
\hline
70418-01-01-00	&	2002-10-08 12:54	&	\cha\ obs.\ 1  (2708)\,\,\,	\\
\hline
95338-01-02-00	&	2010-07-24 06:46	&	\cha\ obs.\ 2 (11815)	\\
\hline
95338-01-03-01	&	2010-07-30 16:50	&	\cha\ obs.\ 3 (11816)	\\
95338-01-03-02	&	2010-07-30 18:36	&	$\cdots$	\\
95338-01-03-03	&	2010-07-30 20:16	&	$\cdots$	\\
95338-01-03-04	&	2010-07-30 21:53	&	$\cdots$	\\
\hline
95338-01-01-04	&	2010-08-01 01:27	&	\cha\ obs.\ 4 (11814)	\\
95338-01-01-03	&	2010-08-01 03:02	&	$\cdots$	\\
95338-01-01-02	&	2010-08-01 04:37	&	$\cdots$	\\
95338-01-01-01	&	2010-08-01 06:11	&	$\cdots$	\\
95338-01-01-05	&	2010-08-01 07:45	&	$\cdots$	\\
\hline
95338-01-03-00	&	2010-08-03 11:30	&	\cha\ obs.\ 5 (11817)	\\
95338-01-03-05	&	2010-08-03 16:36	&	$\cdots$	\\
\hline
95338-01-01-00	&	2010-08-05 15:38	&	\cha\ obs.\ 6 (11818)	\\
95338-01-01-06	&	2010-08-05 17:18	&	$\cdots$	\\
95338-01-01-07	&	2010-08-05 18:57	&	$\cdots$	\\
\hline
\end{tabular}
\label{tab:obs}
\end{center}
\end{table}

To estimate the X-ray luminosity of \gx\ we analyzed \xte/PCA spectra from locations near the HB/NB vertex for each of the three tracks shown in Figure \ref{fig:cd}. For the left panel we used ObsID 95338-01-02-00 (last four orbits, since they were closest to the HB/NB vertex), for the middle panel we used ObsID 40022-01-02-00, and for the right panel we used ObsID 95338-01-03-00 (first orbit). The spectra were extracted from {\tt standard 2} data from Proportional Counter Unit 2, with all layers selected. The spectra were dead-time-corrected, background-subtracted, and a systematic error of 0.6\% was applied. The final spectra were fitted in the 3--40 keV range with XSpec 12.9.0 \citep{ar1996}. We used a simple phenomenological model based on that used for  NS LMXBs by \citet{lireho2009}, consisting of an absorption model ({\tt tbabs}) a diskblack body ({\tt diskbb}), a blackbody ({\tt bbody}), a power law ({\tt pegpwrlw}) and two edges ({\tt edge}).  The spectra were fit simultaneously, with the disk blackbody and blackbody parameters and power-law normalization allowed to vary independently. The \nh\ was tied between the three observations: (5.1$\pm$0.2)$\times10^{22}$ atoms\,cm$^{-2}$, while the power-law index was fixed at 2.5. After extrapolating the best fit model to 0.1--100 keV and assuming a distance of 7 kpc \citep{bashch1999} we find unabsorbed luminosities of 1.2--1.3$\times10^{38}$ erg\,s$^{-1}$ ($\sim$0.7--0.8 \ledd). Such  luminosities are consistent with the high ionization parameters ($\xi$) required to explain the observed Fe \textsc{xxv}--Fe \textsc{xxvi} ratios: log $\xi>4$ \citep{uemuya2004,majodi2014}. 

No \xte/PCA observations were made at the time of the near-infrared observations of \citet{bashch1999}, the {\it ASCA} observation of \citet{ueasya2001}, or the eight  \xmm\ observations analyzed by \citet{sipaoo2002} and \citet{disibo2012}.  It is therefore not possible to map these observations onto the CD tracks that are shown in Figure \ref{fig:cd}. We note that \gx\ data from the \xte\ All-Sky Monitor, which have a better coverage of the source, are not of sufficient quality to construct useful CDs or HIDs.

\subsection{\cha}\label{sec:cha}

\gx\ has been observed seven times with the High Energy Transmission Grating Spectrometer \citep[HETGS;][]{cahuda2000} on board the \cha\ {\it X-ray Observatory};  six of the observations had simultaneous \xte\ coverage (see Table \ref{tab:obs}). \citet{majodi2014} present an analysis of all \cha/HETGS observations. We refer to their Table 1 for a log of the \cha/HETGS observations. \citet{majodi2014} find prominent absorption lines in all seven observations, many of which (Fe \textsc{xxvi}, Fe \textsc{xxv}, S \textsc{xvi} and Si \textsc{xiv}) indicate a strong outflow. The maximum outflow velocities reported for these lines  (for observation 3 in \citealt{majodi2014}) range from  $\sim$400 \vel\ (Fe \textsc{xxv}) to $\sim$800 \vel\ (Fe \textsc{xxvi}). In Figure \ref{fig:cd} we show where wind detections are located along the Z track of \gx. Due to the secular motion of \gx\ in its CD, not all \xte\ observations corresponding to the \cha\ observations were part of  the same CD track. The \xte\ data for \cha\ observations 1--3 are best matched with the track in the left panel of Figure \ref{fig:cd} (group C from F15). They trace out the upper part of the NB, the NB/HB vertex, and (perhaps) part of the lower HB. The \xte\ data for  \cha\ observations 4--6 were included in the track for group E by F15, and are highlighted in red in the right panel of Figure \ref{fig:cd}.  These observations trace out the full HB and NB (down to the NB/FB vertex). 

\subsection{VLA}\label{sec:vla}

\gx\ has been observed with the VLA on several occasions. Here we discuss the two VLA (6 cm) observations presented in \citet{howiru2004}, since these had simultaneous \xte\ coverage (see Table \ref{tab:obs}). We refer to that paper for the details on the radio data analysis. The two VLA observations were made $\sim$4 days apart. The \xte\ data for the two VLA observations were included in the track for group D by F15 and they are highlighted in the middle panel of Figure \ref{fig:cd}. During the first VLA observation (shown in green), when the source had a flux density of $\sim$0.25 mJy, the source was on the FB and at NB/FB vertex. The source was much brighter in the radio in the second VLA observation (shown in orange), with flux densities of 1.3--7.2 mJy, and was likely producing a strong jet. In that observation \gx\ was located on the upper NB and at the NB/HB vertex.





\subsection{{\it ASCA} and \xmm}

As mentioned earlier, signatures of a warm absorber were also seen in all {\it ASCA} and \xmm\ observations of \gx. The \xmm\ spectra suggest outflow velocities up to 3700 \vel\ \citep{disibo2012}, which is considerably larger than the velocities found with \cha\ \citep{majodi2014}. However, \citet{disibo2012} point out that it is plausible that the observed blueshifts are overestimated compared to the real ones, due to residual calibration uncertainties in the EPIC pn camera of \xmm. A detailed analysis of a single EPIC pn timing mode observation by \citet{pisadi2014} resulted in the detection of only one possibly blueshifted absorption line: a Fe \textsc{xxvi} K$_\alpha$, with a suggested blueshift of $\sim1500\pm300$ \vel.

\section{Results}\label{sec:results}

As we have discussed in Section \ref{sec:cha} and  shown in Figure \ref{fig:cd}, the six \cha\ observations of \gx\ for which we have spectral state information all  showed indications for disk winds and were carried out when the source was on its HB or NB. This is also the part of the Z track on which the radio luminosity of \gx, and presumably its jet production as well, is strongest (Section \ref{sec:vla}).  Although the \cha\ and VLA observations were not performed simultaneously, they seem to suggest that when \gx\ is on the HB/NB it can produce disk winds as well as radio jets. Radio emission is much weaker on the FB and it is not clear whether the radio emission on the FB is from a compact jet or from previously launched ejecta. The \cha\ observations do not provide information on the presence of disk winds on the FB.

In the \xte\ data set analyzed by F15, \gx\ can be found on the FB $\sim$30\% of the time. None of the six \cha\ observations listed in Table \ref{tab:obs} were made when the source was on the FB. However, given the fraction of time that \gx\ spends on the FB, it is likely that at least some of the {\it ASCA} and \xmm\ observations (nine in total) were done when \gx\ was on the FB. Several of the \xmm\ light curves presented in \citet{disibo2012} show substantial flaring, suggesting that this indeed is the case. This, in turn, indicates that the warm absorber in \gx\ is present along its entire Z track, although its outflow properties on the FB are uncertain.

\section{Other luminous X-ray binaries}\label{sec:other}

Prompted by our findings in \gx, we searched the literature for indications of simultaneous jets and disk winds in other LMXBs.  These indications can either come from winds and jets being associated with the same spectral state (like in \gx) and/or jets and winds being detected close in time. Given the high luminosity of \gx, our search for other cases focused on, but was not limited to, other LMXBs with near- or super-Eddington luminosities. 

\subsection{NS LMXBs}

We first consider the other Z sources. While all six of the classical Z sources (Sco X-1, GX 17+2, GX 340+0, GX 349+2, Cyg X-2, and GX 5--1) have shown evidence for radio jets \citep{fehe2000,mife2006}, only one has shown signatures of disk winds in its X-ray spectra: GX 340+0. \citet{miraca2016} report on the detection (with \cha) of absorption features in the Fe K band of this source. The strongest of these features, when fitted with photoionization models, implies a fast disk wind with an outflow velocity of $\sim1.2\times10^{4}$ \vel.  \citet{miraca2016} do not state on what part of the Z track the wind was detected. We used the {\tt  aglc} package for the Interactive Spectral Interpretation System \citep{hode2000} to create an HID from the four archival \cha/HETGS observations of GX 340+0. The results of a comparison with a more complete HID constructed from \xte\ data (using similar energy bands) are ambiguous, given the differences in the shapes of the two HIDs (likely due to differences in the detector responses within the chosen energy bands). However, based on the intermediate hardness of the \cha\ observation in which the disk wind was detected, our best guess is that the observation was done when the source was on the NB. \citet{ooleva1994} reported on simultaneous X-ray and radio observations of GX 340+0, during which the source traced out the NB and FB. The highest radio fluxes were detected when the source was on the NB. This could indicate that, like \gx, when GX 340+0 is on the NB it can produce radio jets and/or disk winds, although this interpretation strongly relies on our best guess of the source state during the disk wind detection with \cha.

While none of the other classical Z sources have shown evidence for disk winds in their X-ray spectra \citep{uemimu2005,camiho2009,schuji2009,lufa2014},  signatures of a disk wind were seen in near-infrared spectra of Sco X-1.  \citet{bashch1999} detected a P Cygni profile in the Br$\gamma$ line, indicating an outflow velocity of $\sim$2600 \vel. We note that \citet{mamuca2015} did not see a P Cygni profile for the Br$\gamma$ line in their observations, possibly indicating that the disk wind in Sco X-1 is variable. It is not clear on which Z source branches these near-infrared observations took place. However, given the fact that the compact radio core of Sco X-1 is seemingly always detected \citep{fogebr2001a,fogebr2001b,parama2005}, the detection of disk wind signatures  in the near-infrared suggests that Sco X-1 can (at least occasionally) produce simultaneous jets and disk winds, regardless of the branch it was on during the  observations of \citet{bashch1999}. 

Cir X-1 is a NS LMXB that shows clear Z source behavior at its highest luminosities \citep[]{shbrle1999} and turns into an atoll source at lower luminosities (F15). \citet{brsh2000} and \citet{scbr2002} observed the source twice with \cha\ (close to orbital phase 0.0) in 2000, i.e.\ when Cir X-1 was still exhibiting Z source behavior, and detected several P Cygni lines that indicate a strong disk wind with inferred outflow velocities of $\sim$200--1900 \vel. Both of these \cha\ observations were carried out simultaneously with \xte\ observations. The CD and HID of the first \xte\ data set can be seen in panel (C) of Figure 9 in F15. A full Cyg-like Z track is traced out, showing a dipping-FB, NB, HB, and a strong HB-upturn (see F15 for details). While the wind is present along the entire track, there is significant evolution in the profile of several of the strongest P Cygni lines (see Figure 7 in \citealt{scbr2002}). The second \cha\ observation was likely more strongly affected by absorption (given the very high soft color values of the accompanying \xte\ data) and as a result the Z source branches could not be clearly identified in the \xte\ data. \citet{tufetz2008} present an analysis of radio data taken between 1996 and 2006. Although radio flares from Cir X-1 experienced a long-term lull during that period \citep{arfeni2013}, the source was still active in the radio, with flux densities between  a few mJy and a few tens of mJy. Radio  emission from the core generally appeared to be present at all orbital phases, but was highly variable and its relation to position along the Z track is not well understood \citep{sotufe2009}. While the \cha\ and radio data therefore support the possibility of jets and disk winds  being present at the same time during the Z source phase of Cir X-1, especially given the presence of a disk wind along the entire Z track of the first \cha\ observation in 2000, the case is not as clear as those for \gx\ and Sco X-1. We note that \cha\ observations in 2005, when the source had  become more similar to a bright atoll source (based on an inspection of archival \xte\ data; see also F15), no longer revealed P Cygni lines and no signs of disk outflows were seen at these lower luminosities \citep{sckaga2008}.

\subsection{BH LMXBs}

As mentioned in Section \ref{sec:intro}, evidence for simultaneous jets and disk winds was already reported for the BH LMXB GRS 1915+105 by 
\citet{lerere2002} and \citet{nele2009}. During the single observation in which this was found \citep[listed as ``H1'' in][]{nele2009} GRS 1915+105 was in the spectrally hard $\chi$ state \citep{beklme2000}. A fit with a simple phenomenological model to \xte/PCA spectra taken at the same time yields an unabsorbed 0.01--100 keV flux of 5.2$\times10^{-8}$ \flux. Using the parallax distance (8.6 kpc) and revised mass estimate (12.4 $M_\odot$) from \citet{remcst2014}, this gives a luminosity of $\sim$0.3 \ledd. Although this is well below \ledd, it is higher than the typical hard-state peak luminosities of transient BH LMXBs \citep{dufeko2010}, suggesting that a high luminosity is fundamentally important to produce winds in BH hard states (see also Section \ref{sec:disc}).

V404 Cyg showed an outburst in 2015, during which the source frequently approached and perhaps even exceeded \ledd\ \citep{jewiho2016,ranaag2016}. The source was observed twice with \cha\ during this outburst. Blueshifted emission lines with velocities up to $\sim$900 \vel\ were detected and in the highest flux phases lines with P Cygni profiles were seen (\citealt{kimira2015}; A.\ L.\ King 2016, private communication). The intrinsic luminosities during these \cha\  observations, while probably still sub-Eddington, were likely well above the 0.005--0.05 \ledd\ luminosities implied by the (non-absorption-corrected) 2--10 keV fluxes.  During the same outburst \citet{mucama2016} observed P Cygni profiles in various optical H and He emission lines, indicating a disk wind with terminal outflow velocities in the range of $\sim$1500--3000 \vel.  V404 Cyg was active in the radio at the time of these optical and X-ray observations \citep{trnity2015,tsaoas2015}, which suggests that the source was producing disk winds and jets at the same time, as already pointed out by \citet{kimira2015}. We note that {\it Fermi}/GBM observations indicate that V404 Cyg remained in the hard state throughout its 2015 outburst \citep{jewiho2016}, but we note there was often strong variability within observations. Simultaneous jet and disk wind production likely also occurred during the 1989 outburst of the source; \citet{cachjo1991} detected P Cygni profiles in optical H and He emission lines when the source was still bright in the radio with a flat spectrum \citep{hahj1992}.

\citet{necofe2014} report on simultaneous \cha/HETGS and Australia Telescope Compact Array radio observations during an outburst of 4U 1630$-$47. In 2012 January, when 4U 1630$-$47 was in a spectrally soft state ($\gtrsim$0.1 \ledd, for a 10 $M_\odot$ BH), they detected clear X-ray signatures of a disk wind (with speeds of $\sim$200--500 \vel) while the source was also detected in the radio. However, the radio emission was optically thin at that point, suggesting that it was due to jet--ISM interaction far from the BH, rather than from a jet that was being launched at the same time as the disk wind.

Finally, although it is not an LMXB, it is worth mentioning the high-mass X-ray binary SS433 \citep{lomaca2006}, which likely harbors a BH. This famous jet source (see \citealt{fa2004} for a review) is a supercritical accretor and shows a dense equatorial wind from the accretion disk \citep{gimcri2002} with a velocity up to $\sim$1300 \vel, which feeds into a circumbinary ring \citep{blbosc2008}.

\section{Discussion}\label{sec:disc}

The above results show that jets and disk winds in LMXBs are not necessarily mutually exclusive.  The common residence of disk winds and
radio emission in (the spectrally hard states of) various luminous LMXBs, and in the HMXB SS433,  suggests
that the presence of disk winds does not necessarily result in the suppression of radio jets, as was suggested by \citet{nele2009} for GRS 1915+105. 
However, there is likely a time scale over which such suppression would happen, e.g.\ the time it takes for a mass-energy deficit in the disk to propagate inward to the jet launching region. In such a scenario, occasional simultaneous detections of jets and disk winds are expected to happen, even if jets and disk winds are largely mutually exclusive. With regard to this, it is also important to stress that detecting radio emission at a given moment does not mean a jet is being produced right at that moment. In the case of short-lived transient jets, as observed in, e.g., GRS 1915+105 \citep{klfepo2002},  we may be detecting radio emission from jet ejecta, rather than from a compact jet. Radio spectral indices are needed to distinguish between the two types of radio emission. Similarly, time-averaged X-ray spectra with signatures of disk winds do not necessarily mean that disk winds are present for the entire duration of such an observation \citep[e.g.,][]{nerele2011}. Long  observations (hours to days) with sufficient time resolution (minutes or seconds) are needed to both study the transient nature of winds and jets, and to constrain the time scale on which jet suppression by winds possibly occurs. Sources in luminous steady hard states would be the preferred targets for such observations.

It is also important to point out that the non-detection of narrow absorption lines does not automatically indicate the absence of disk winds. Strong photoionization can fully ionize a wind, resulting in the disappearance of absorption lines \citep{dimimi2014,dibo2015}.

In the sources studied by \citet{pofebe2012,pomufe2014,pobimu2015} winds are predominantly observed in spectrally soft states with luminosities up to a few tens of percent of \ledd.   These winds are thought to be thermally driven as the result of Compton heating \citep{bemcsh1983}. The general absence of winds in spectrally hard states with similar luminosities might be the result of less efficient Compton heating due to, for example, a large fraction of the X-rays being beamed away from the outer disk \citep{mabepo2001}, or a disk geometry that is less favorable to irradiation \citep{pofebe2012}. For the three systems discussed in Section \ref{sec:results} for which we have state information, we found that the simultaneous jets/winds were observed in the sources' spectrally hardest states (i.e., with the strongest non-thermal contribution): the HB/NB for \gx\ and the hard state for GRS 1915+105 and V404 Cyg. These states have historically only been associated with radio jets. The reason for the presence of disk winds in these `jet states' is probably the relatively high X-ray luminosity of the sources, ranging from $\sim$0.3 \ledd\ in GRS 1915+105 to near- or super-\ledd\ in V404 Cyg. This is higher than the peak hard-state luminosities observed in other high-inclination BH or NS LMXBs \citep{pofebe2012,pomufe2014,pobimu2015}. 
The simplest explanation for winds in these spectrally hard states is that the high luminosities compensate for intrinsically less efficient Compton heating of the accretion disk in such states. The near- or super-Eddington luminosities in V404 Cyg, \gx, Sco X-1, and Cir X-1 further suggest that the winds in those systems may be partially radiatively driven, since radiation pressure on electrons can become an important factor \citep{prka2002}.

Winds have also been suggested as a by-product of radiatively inefficient accretion flows \citep{blbe1999}, which are thought to exist in spectrally hard states \citep{nayi1994,yuna2014}. There is observational evidence \citep{wanoma2013} and theoretical support \citep{yagana2015} for this. These winds are not thermally or radiatively driven, but are rather driven by centrifugal forces and magnetic pressure gradients. It is not clear why they would only be seen in higher luminosity hard states, or whether they should be observationally distinguishable from the winds produced in softer states.

The P Cygni Br$\gamma$ line detected in the near-infrared spectrum of \gx\ indicates an outflow speed for the disk wind of $\sim$2400 \vel \citep{bashch1999}. This is substantially higher than the maximum outflow speed inferred from the \cha\ spectra ($\sim$800 \vel). Since the Br$\gamma$ line is produced further out in the disk wind than the X-ray absorption lines, this could indicate that the wind is still being accelerated at large radii (assuming the lines are all produced by the same wind). However, simultaneous X-ray and near-infrared spectra of \gx\ (and other LMXBs) are needed to verify this.

Finally, we note that the simultaneous presence of jets and disk winds is not unique to luminous X-ray binaries and appears to be fairly common in radio-loud active galactic nuclei (AGNs), which are characterized by their strong jet emission. \citet{totamu2014} studied 26 of these systems and concluded that in 50$\pm$20\% of the radio-loud AGN (fast) winds are present. Unlike the winds in LMXBs, these winds are not preferentially equatorial and have a large opening angle. Radio-loud AGNs also span a wider range in fractional \ledd\  \citep[$\sim10^{-2.5}-1$;][]{sistla2007} than the X-ray binary systems in which we see indications of simultaneous jets and winds.  The fact that radio-loud AGNs can also produce winds at lower \ledd\ fractions is possibly due to more efficient UV line driving \citep{prka2004}.







\acknowledgments

We thank the referee for his/her constructive comments. This research has made use of data obtained from the High Energy Astrophysics Science Archive Research Center (HEASARC), provided by NASA's Goddard Space Flight Center, and data obtained from the \cha\ Data Archive. J.N. acknowledges support from NASA through the Hubble Postdoctoral Fellowship program, grant HST-HF2-51343.001-A. Finally, we thank David Huenemoerder for reprocessing the \cha\ data of GX 340+0.


\end{document}